\documentclass[aps,twocolumn,showpacs,preprintnumbers,amsmath,amssymb,nofootinbib,superscriptaddress,showkeys]{revtex4}

\usepackage{epsfig}
\usepackage{graphicx}
\usepackage{mathptmx}      % use Times fonts if available on your TeX system
\usepackage{latexsym}

\def\slashchar#1{\setbox0=\hbox{$#1$}
   \dimen0=\wd0 \setbox1=\hbox{/} \dimen1=\wd1
   \ifdim\dimen0>\dimen1 \rlap{\hbox to \dimen0{\hfil/\hfil}} #1
   \else  \rlap{\hbox to \dimen1{\hfil$#1$\hfil}} / \fi}

\begin{document}

\title{Reply to "Comment on 'Systematics of radial and angular-momentum Regge trajectories of light non-strange $q\bar{q}$-states' "
\footnote{Supported by the Deutsche Forschungsgemeinschaft DFG through the Collaborative Research Center ``The Low-Energy Frontier of the Standard Model" (SFB 1044), by MICINN of Spain (FPA2010-16802, FPA2010-16696, FIS2011-24149) and Consolider-Ingenio   2010 Programme CPAN (CSD2007-00042), by Junta de Andaluc\'{\i}a (FQM 101, FQM 437, FQM225 and  FQM022) and by the Polish Science and Higher Education, grant N~N202 263438, and National Science Centre, grant DEC-2011/01/D/ST2/00772.}}

\author{Pere Masjuan} \email{masjuan@kph.uni-mainz.de}
\affiliation{Institut f\"ur Kernphysik, Johannes Gutenberg-Universit\"at, D-55099 Mainz, Germany.}

\author{Enrique Ruiz Arriola}\email{earriola@ugr.es}
\affiliation{Departamento de F\'{\i}sica At\'omica, Molecular y Nuclear and Instituto Carlos I de F{\'\i}sica Te\'orica y Computacional \\
Universidad de Granada, E-18071 Granada, Spain.}

\author{Wojciech Broniowski} \email{Wojciech.Broniowski@ifj.edu.pl}
\affiliation{CNRS, URA2306, Institut de Physique Th\'eorique de Saclay,
F-91191 Gif-sur-Yvette, France}
\affiliation{The H. Niewodnicza\'nski Institute of Nuclear Physics,
Polish Academy of Sciences, PL-31342 Krak\'ow, Poland} 
\affiliation{Institute of Physics, Jan Kochanowski University, PL-25406~Kielce, Poland}

\date{\today}

\begin{abstract}

In his Comment, D.~Bugg argues against our usage of the PDG collection of light non-strange states together with the half-width rule to analyze the linearity of radial and angular-moment Regge trajectories in the large-$N_c$ limit. After taking into account his observations on our choice of data, the radial Regge trajectories are again analyzed. We still find that our conclusion on the lack of universality between radial- and angular-momentum Regge trajectories is valid.

\end{abstract}



\maketitle

\section{Introduction}

In our paper \cite{Masjuan:2012gc} we reanalyze the radial $(n)$ and angular-momentum $(J)$ Regge trajectories for all light-quark states with baryon number zero listed in the 2011 edition of the Particle Data Tables. The parameters of the trajectories were obtained with linear regression, with weight of each resonance inversely proportional to its half-width squared, what we called the \textit{half-width rule} \cite{RuizArriola:2010fj,Arriola:2011en,Masjuan:2012yw,Masjuan:2012sk,Arriola:2012vk}. We argued in \cite{Masjuan:2012gc} that the half-width rule fully complies to the large-$N_c$ viewpoint and actually suggests an interesting interpretation: the Regge-fitted masses are considered to be the leading-$N_c$ contribution to the mass of the resonance. This incorporates a desirable flexibility as to what should the Regge fit be compared to.
 
In his Comment~\cite{Bugg:2012yt}, D.~Bugg, making use of the Crystal Barrel determinations, argues against our usage of the half-width rule to analyze the  PDG collection of  light non-strange $q\bar{q}$-states in order to study the radial Regge trajectories. From a theoretical point of view, he concludes in his Comment that the $N_c$ world is different from $N_c \rightarrow \infty$. From a phenomenological point of view, he disagrees in our selection of states in the fit.

It should be pointed out that the Comment and our work not only are made on different levels but actually have different scopes.  In the Comment a particular parameterization is taken to assess the accuracy of the determination of resonance masses. No attempt to discuss any possible systematic errors is undertaken, thus it is unclear whether such an accuracy turns out
to be a compromise between a bias in the choice of the particular method or the attempt to make such a choice to comply with the data.

In contrast, our work provides an attempt to determine the validity of the linear Regge formula for the leading large-$N_c$ contribution to the
mass of the mesonic state for both the radial and the angular-momentum trajectories. However, most states entering the validation of the
formula are unstable states, i.e. resonances, which on a rigorous basis are characterized as poles of the $S$-matrix on the second Riemann
sheet of the complex plane of the Mandelstam s-variable. This provides a unique and process-independent definition, whereas the corresponding complex residue reflects the coupling to the background process where the resonance is generated.  We understand that in any quantum field theory, and in QCD in particular, the positions of such poles are a well defined concept. However, complex energies cannot be measured and an analytic continuation to the complex plane is required. If the amplitude on the real axis is just approximated or a model is used, the analytic continuation amplifies the uncertainty in the resonance parameters. This, of course, requires providing a definition on what should be called the ``experimental'' resonance mass. Although there are a few definitions, any of them reflects both the distance to
the real axis as well as the influence of the background. In other words, different models and/or definitions provide different numbers and comparisons require to take this into account.  

From a purely theoretical point of view there is no obvious reason why the Regge formula for mesons $M^2 = a n + b J + c$ should be exact in QCD and the puzzling fact that it works so nicely demands an explanation. But it is far from obvious that the Breit-Wigner definition or a pole definition should reproduce the Regge formula any better than other choices. It is this kind of a puzzle that motivated our research. 

As argued in our article, a possible model-independent way of approaching the problem from the point of view of QCD is by invoking the large-$N_c$ expansion, where resonances become stable since the ratio of width to mass is $1/N_c$ suppressed. We quoted the value deduced from the PDG for this ratio, namely $0.12(8)$~\cite{Masjuan:2012gc}, which follows from using all the listed resonances and which
suggests the goodness of the approach for the present discussion.  Further, the mass shift is also $1/N_c$ suppressed, which is parametrically the same as the width, such that we take half of the width as a crude (and possibly too large) estimate. On the other hand, different methods to define the width, such as the Breit-Wigner, the time delay, the speed plot, etc., would yield different numerical values for the resonance mass that might not exceed this half-width estimate.  This consideration solves the problem of the magnification of errors when going from the real axis to the complex plane.  While this may be a too conservative estimate concerning the mass shift, it certainly provides a credible upper bound
buttressing the half-width rule. In spite of this crude estimate, the globality of the analysis provides some accurate statements which were presented as the main result of our article.  This was the framework of our motivation and conclusions. We have never pretended to go beyond that.

The reader should also notice that we have considered other fit functions inspired by other theoretical frameworks (such as a linear
n-dependence of the masses) and our choice is the one favored by the quality of the fits within the generous half-width rule uncertainties.

\begin{figure*}[htpb]
\begin{center}
\includegraphics[width=8cm]{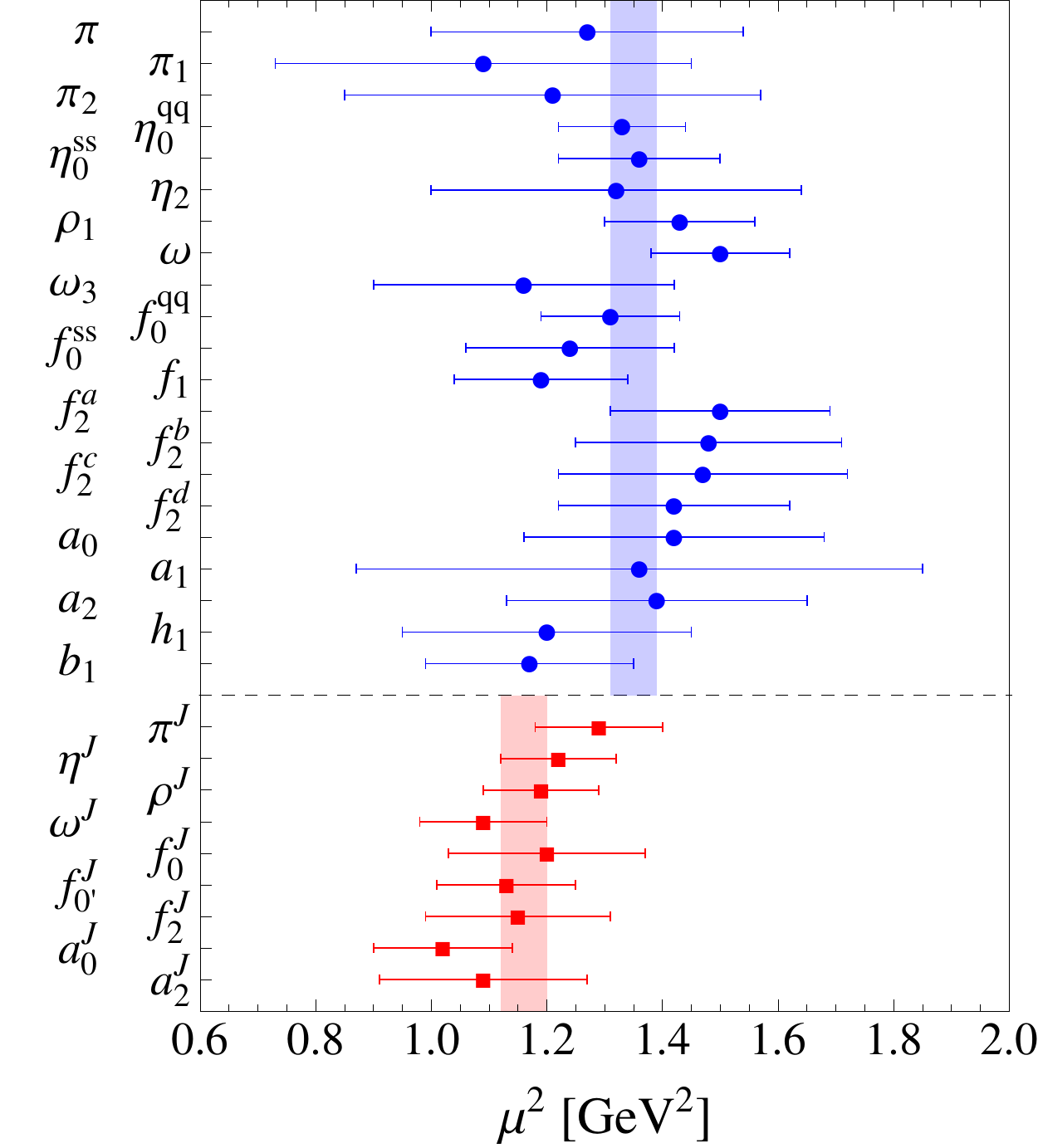}
\includegraphics[width=8cm]{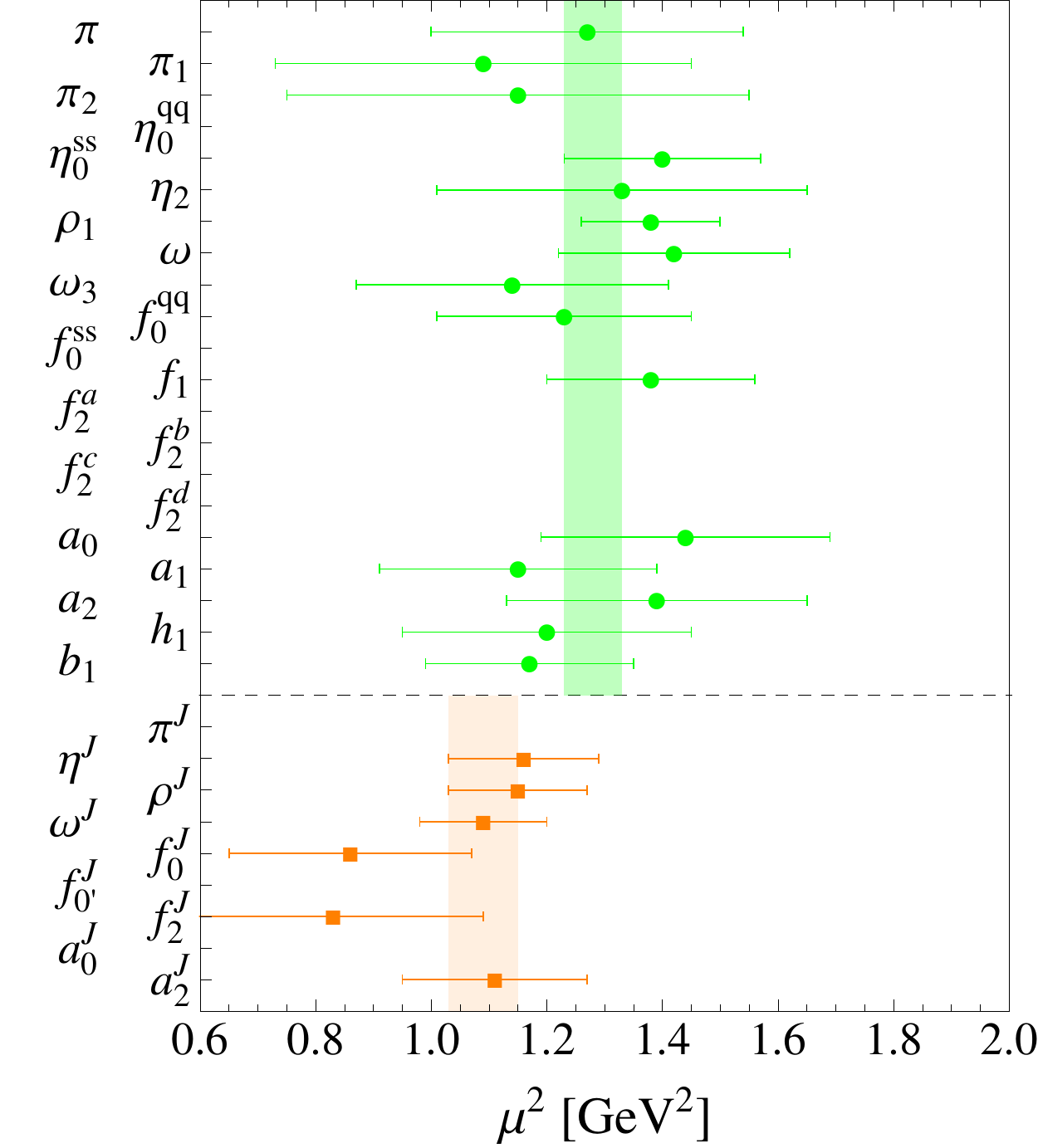}
\caption{{\bf Left}: $(n,M^2)$ and $(J,M^2)$ slope results for the trajectories considered in Ref.~\cite{Masjuan:2012gc}; {\bf Right} $(n,M^2)$ and $(J,M^2)$ slope results for the trajectories considered in the Comment with three or more states obtains with the half-width rule. The horizontal dashed line separates the radial slopes (circles) from the angular-momentum slopes (squares). Individual errors are estimated from the $\chi^2$ fits to the corresponding trajectories. The bands correspond to the weighted averages of the radial (upper band) and the angular-momentum (lower band).}
\label{plot}
\end{center}
\end{figure*}

From a more phenomenological point of view, all the information on the states collected in our publications are extracted from the Particle Data Group Tables. While this may be unsatisfactory for the author of the Comment for certain states which might be slightly misplaced, within the half-width uncertainty these changes do not modify our global picture. The question on whether the Crystal Ball measurements are underrepresented in the PDG compilation, while interesting in its own, was naturally not within the scope of our study. In particular, we wanted to test the hypothesis of the universality of the angular-momentum and radial Regge trajectories and for that our usage of the data needs to be "global". In this regard the Particle Data Tables provide us with the best compilation of the data. 

We have two categories of radial Regge trajectories. In the first category we include the trajectories with three or more states, all
of them considered by the Particle Data Group as a well determined resonance. We establish a final averaged value for the slope of the
radial Regge trajectory considering only this set of trajectories. In the second category we include all the trajectories with only two states, or states not confirmed by the Particle Data Group, although commented and included in our analysis for completeness. This category is not taken into account for the final compilation. We understand that a line can always go through two points but this does not confirm or refute the linearity of the Regge trajectories. In this respect, trajectories with large slope error such as the $a_4$ pointed out in the Comment produce null impact in our averaged result.

Considering our published work together with the Comment, after all the scrutiny and detailed discussion, we do not see such a difference
and such an amount of defects in the choice of data as claimed by the author of the Comment. In fact, we agree on
the $f_1, f_3, b_1, b_3, h_1, h_3, \omega_3, a_2$, $a_4, \pi,\pi_2, \rho, \rho_3, \eta, \eta_2$ families (accounting for more than 50
states) and only disagree in the classification of a few states from the $\omega, f_0, a_0, a_1, f_2$ families (less than 10 states). Taken
into account that the $\pi_1$, $\phi$, $\eta$' and $a_3$ families do not appear in the Comment (there are around 25 states not considered 
in it), we indeed find quite a good agreement in the selection of the data in both approaches. We notice that our inspection of states also
includes $\eta_4, \rho_4, \rho_5, \pi_4, a_4,a_6, f_6$, $\omega_4$ and $\omega_5$ in the angular-momentum Regge trajectory analysis.

In this regard, we repeat our fits with the subset of states provided in the Comment (accepting the extreme case that all the criticism raised there is correct and that our selection of data were wrong), keeping the half-width rule and only trajectories with three or more states (following the previous argumentations). In this new scenario we would have fewer radial trajectories as shown in Fig.~\ref{plot}. Indeed, for ease of comparison with our previous published results (shown in the left panel of Fig.~\ref{plot}), we keep the same labeling in the right panel of Fig.~\ref{plot} but with empty result for the trajectories rejected in the Comment. The new weighted averaged slope for radial Regge trajectories would be 
\begin{equation}
\mu^2=1.28(5) \textrm{GeV}^2\, ,
\end{equation}
\noindent
and for the angular-momentum trajectories
\begin{equation}
\beta^2=1.09(6) \textrm{GeV}^2\, ,
\end{equation}
to be compared with $\mu^2=1.35(4)\textrm{GeV}^2$ and $\beta^2=1.16(4)\textrm{GeV}^2$ from Ref.~\cite{Masjuan:2012gc}.
We notice that, although the averaged slopes are a bit smaller than in our publication (compatible within errors) the universality of radial and angular-momentum Regge trajectories is still not fulfilled at the level of $2.4$ standard deviations (compared to the $3.4$ standard deviation found previously), so our conclusion that no strict universality of slopes occurs in the light non-strange meson spectra is still valid.

In summary, we believe that both analyses do not oppose necessarily, since the scope is different; we see no obvious reason why the
Breit-Wigner masses obtained from the Crystal Barrel data should be directly compared to QCD-inspired models. We admit an existing
source of systematic error within the large-$N_c$ approach motivated by our attempt to estimate the slopes within that leading order. We
never tried to replace any previous experimental analyses, but rather, to provide some insight taking into account the 1/$N_c$ theoretical
errors.

\end{document}